\documentclass{iau_FM}
\usepackage{graphicx}

\title[Black hole axis perturbations] %% give here short title %%
{Probing black hole axis perturbations using low axial ratio radio galaxies} %% give here the full title %%

\author[Lakshmi Saripalli]   %% give here short author list %%
{Lakshmi Saripalli$^1$}
\affiliation{$^1$Raman Research Institute, Sadashivanagar, Bangalore, India  \\ email: {\tt lsaripal@rri.res.in}} 
\pubyear{2018}
\jname{Astronomy in Focus, Volume 1} 
\editors{Volker Beckmann \& Claudio Ricci, eds.}
\begin{document}
\maketitle

\begin{abstract}
With the unique advantage that radio galaxies have in delineating the central supermassive black hole axis along their radio axes they make useful probes for unseen central engine processes. In this paper I bring to attention the remarkable power of low axial ratio radio galaxy samples in opening up for examination this hard-to-probe regime of active galaxies. I present results from our recent EVLA observations of a large radio source sample originally selected on the basis of shapes. The observations have revealed a plethora of radio sources that potentially show signatures of axis changes of various types suggesting such samples as effective means for probing black hole axis perturbations and hence such samples to be effective means in searches for binary black hole pairs in radio galaxies.
%% add here a maximum of 10 keywords, to be taken form the file <Keywords.tex>
\end{abstract}

\firstsection % if your document starts with a section,
\section{Introduction}

Already in the early years when radio galaxies were being imaged and their basic twin-lobe morphologies recognised and studied
occurrence of deviant structures as in S, Z and X-shaped radio galaxies (XRGs) were noted and models examined as to how they developed 
their morphologies (e.g. \cite[Ekers (1978)]{Ekers78}). Over the years several such sources were found, serendipitously, and their morphologies of
striking inversion symmetry were mostly modelled as arising due to jet axis precession (e.g. \cite[Begelman et al. (1980)]{Begelman_etal80}, 
\cite[Gower et al. (1982)]{Gower_etal82}, \cite[Cox et al. (1991)]{Cox_etal91}). 
XRGs came in for separate attention given their structures, which
exhibited two lobe pairs vastly separated in position angle. Initial models included the backflow deflection model 
(\cite[Leahy \& Williams (1984)]{LeahyWilliams84}) in which, as backflows in powerful lobes flow towards the central regions
they get deflected in opposite directions in the density gradients in an asymmetric host atmosphere and thereby create the transverse 'wings'. Also suggested 
was a model where a black hole associated with a radio source undergoes an axis precession and renews its activity in the new direction
with backflows from the new lobes energising the older relic lobes (also \cite[Leahy \& Parma (1992)]{LeahyParma92}). In later years in the backdrop of growing 
interest in binary black hole systems and gravitational waves \cite[Merritt \& Ekers (2002)]{MerrittEkers02} suggested what became known as spin-flip model where
the black hole associated with a radio source undergoes a rapid large-angle spin axis flip and continues its activity in the new direction. Over time these two 
quite contrasting models, the backflow deflection model and the rapid spin-flip model became the main contending scenarios for XRG formation.

Despite both XRG formation models enjoying support in observational and simulation realms, e.g. requirement within the backflow deflection model of asymmetry
and main lobes to be of FR-II type are both met (\cite[Capetti et al. (2002)]{Capetti_etal02}, \cite[Saripalli \& Subrahmanyan (2009)]{SaripalliSubrahmanyan09}, \cite[Hodges-Kluck et al. (2010)]{Hodges-Kluck_etal10}, \cite[Hodges-Kluck et al. (2011)]{Hodges-Kluck_etal11})
and separately, black hole spin-flips are clearly found to occur in simulations of binary black hole systems (\cite[Campanelli et al. (2007)]{Campanelli_etal07}), 
neither model is devoid of serious drawbacks. For example, subsonically advancing wings in XRGs that have extents same as or larger than the 
supersonically advancing main lobes are a problem within the backflow scenario, whereas prevalence of relic radio sources in the form of wings in XRGs 
when relics are known to be rare
is difficult to understand within the spin-flip model (besides central emission gaps, smooth continuity between a wing and associated main lobe, 
main lobes needing to be of FR-II type, and wings forming along the host minor axis (\cite[Saripalli \& Subrahmanyan (2009)]{SaripalliSubrahmanyan09} and 
\cite[Saripalli \& Roberts (2018)]{SaripalliRoberts18}). 

It was with the motivation of obtaining clarity in this regard that we embarked on EVLA observations (Roberts et al, 2018) of a large sample of 100 candidate 
XRGs (\cite[Cheung (2007)]{Cheung07}), originally selected on the basis of their shapes from the FIRST survey. 
The multi-band and multi-array total intensity and polarization continuum observations were carried out over a year and (including our
analysis of archival VLA data of a large subset of 55 sources) resulted in 95 sources being 
imaged in multiple frequencies and resolutions. This allowed a thorough characterisation of the sample sources  (\cite[Saripalli \& Roberts (2018)]{SaripalliRoberts18}), an analysis that
draws attention to the immense opportunity that low axial ratio radio sources provide us for probing black hole axis behaviour and perturbations to black hole axes 
in radio galaxies. 

\section {Source morphologies in low axial ratio radio source samples} 
Our EVLA observations of a sample selected on the basis of low axial ratios has revealed the sample to be dominated by inversion symmetric radio morphologies. Strikingly, the inversion symmetry is bimodal with off-axis emission features originating from two strategic locations, the outer ends of lobes (Outer-deviation sources or O-devs) and 
inner ends of lobes (Inner-deviation sources or I-devs). Remarkably, the more dominant I-devs are invariably of FR-II type. Nearly one-third of I-devs have at least one wing extent 
same or larger than their main lobes. Trends are also seen of longer wings likely associated with lobes having stronger hotspots and of 
sources with fractionally longer wings having physically smaller sizes (also reported in a separate study of XRGs by \cite[Saripalli \& Subrahmanyan (2009)]{SaripalliSubrahmanyan09}). We identify a separate class of sources as XRGs, where there is a central swathe of emission that cannot be clearly linked to any individual lobe. Two-thirds of O-dev sources are either of S- or Z-shape. A large fraction, nearly one-third of sources in the entire sample, exhibit inner-S shape in their main radio axes. Sources with double hotspots in both lobes are seen where the hotspots appear to lie along two separate axes through the core. The imaged sample includes a small subset with evidence of two separate radio sources along separate axes through the core.

\begin{figure}[b]
% \vspace*{-2.0 cm}
\begin{center}
 \includegraphics[width=3.4in]{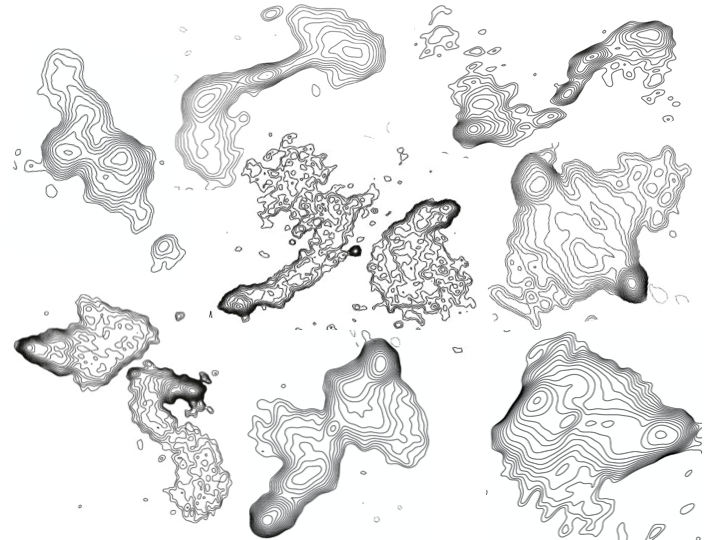} 
% \vspace*{-1.0 cm}
 \caption{Montage of radio galaxy morphological types in the low axial ratio source sample}
   \label{fig1}
\end{center}
\end{figure}

\section {Inner-deviation sources and XRGs}
In the literature the term 'XRG', besides referring to the classic X-structure, includes X-shaped sources with central emission gaps where each of the wings is separately associated with a main lobe. We choose to separate the two types using instead 'XRG' for the classic X-shaped sources and I-dev for the group with emission gaps. The present study allowed us to address the original motivation for the project, that of 
the formation scenario for XRGs. In revealing I-devs to invariably be associated with FR-II morphologies the present work shows that backflows cannot be ignored in the formation of central distortions. However, as pointed out above, there remain areas where both models encounter problems. Combining the main aspects of the two models, namely, black hole spin-flip and backflow deflection in a central asymmetric environment (\cite[Leahy \& Williams (1984)]{LeahyWilliams84}, \cite[Leahy \& Parma (1992)]{LeahyParma92} and suggested for NGC326 by \cite[Hodges-Kluck \& Reynolds (2012)]{Hodges-KluckReynolds12}), we suggested the following scenario for formation of XRGs and I-devs, where after suffering a large-angle axis flip from the host minor axis direction and leaving behind a relic radio source the central engine continues activity in the new direction where any ensuing backflows in the new source would energise the respective parts of the relic after being deflected by density gradients in the asymmetric host halo distribution. With fractionally longer wings likely associated with physically smaller sources as found in the present study, such a combined model has no difficulty in accounting for wings larger than main sources. The central emission gaps seen in several XRGs (called I-devs here) may develop as the source evolves, affected by age and buoyancy effects in the host atmosphere. Interestingly, we find the average linear size of I-devs is more than twice (354 kpc) that of XRGs (151kpc) for the sample we studied.

\section {Low axial ratio radio sources as probes for black hole axis perturbations}
Our EVLA observations of the large sample of low axial ratio radio galaxies have revealed source structures that exhibit a variety of signs of axis change. At least 4\% of radio galaxies in radio source samples are found to likely suffer axis perturbations. Black hole axis precession is a favourite mechanism to explain S- or Z-shaped sources and
also those with double hotspots as in Cygnus-A (\cite[Cox et al. (1991)]{Cox_etal91}), where precession can arise in situations of re-aligning black holes or tilted or warped accretion disks or binary black hole systems, which also give rise to large-angle spin axis-flips (\cite[Natarajan \& Pringle (1998)]{NatarajanPringle98}; \cite[Liu (2004)]{Liu04}; \cite[Begelman et al. (1980)]{Begelman_etal80}; \cite[Campanelli et al. (2007)]{Campanelli_etal07}). Given the variety of conditions in which such processes could occur, large samples of low axial ratio radio sources with the variety of source morphologies such as what we have characterised, which between them would include a range of precession parameters, are ideal for investigating the variety in physical conditions likely involved.  

\section {Implications}
The implications of such a study described here are immense for galaxy merger evolution studies, particularly in the final stages. Efforts in this area have included searches for
dual and binary black holes that allow to identify individual galaxies and to study the conditions that prevail at the particular stage of merger and also study of 
the evolution of the merger from tens of kiloparsec separations to the final stages of binary black holes. The workhorse in this effort has been the strategy of using 
double-peaked emission line galaxies as markers for dual AGN (\cite[Muller-Sanchez et al. (2015)]{Muller-Sanchez_etal15}), although this approach is recognised for the tedious effort involved. A newer technique of using mid-IR selected galaxy merger candidates has recently found several dual AGN candidates, promising to be a more efficient method (\cite[Satyapal et al. (2017)]{Satyapal_etal17}). Even so, binary AGN have been difficult to find. They require high resolution and with radio observations capable of providing the highest resolution through VLBI and with availability of abundant target populations as shown in this study, it appears that
using low axial ratio radio samples may be an effective method. With two-thirds of our sample sources possibly candidates for axis perturbation, low axial ratio radio samples may be a strong search method for binary black holes as well as recoiling black holes.

\section {Conclusions and Future work}
With the present work showing a way to generate large samples of sources with signatures of axis change, low axial ratio radio sources are shown to be potentially effective and useful resource for searching for binary black holes and recoiling black holes as well as in capturing the wide parameter space in study of galaxy mergers into extreme gravity regimes.
Having isolated candidates for perturbed black holes, follow up studies of the host galaxy, radio spectra, VLBI observations, axis precession modelling etc are needed towards examining S-, Z-, XRG and I-dev formation scenarios and towards gaining clues to prevailing conditions as galaxy merger progresses. A few are presently in progress.

\end{document}